\documentclass[a4paper,12pt]{article}
\usepackage[psamsfonts]{amssymb}
\usepackage{amsmath}
\usepackage{epsfig}
\def\be {\begin{equation}}
\def\ee {\end{equation}}
\def\ba {\begin{eqnarray}}
\def\ea {\end{eqnarray}}
\def\nn {\nonumber}
\def\bea{\begin{eqnarray}}
\def\eea{\end{eqnarray}}
\def\bi {\begin{itemize}}
\def\ei {\end{itemize}}
\begin{document}

\title{\bf \large {Non-minimal coupling of the phantom field \\and cosmic acceleration}}
\author{\normalsize{M. R. Setare$^{1,2}$\thanks{%
E-mail: rezakord@ipm.ir}  \, and\, Elias C. Vagenas$^{3}$\thanks{%
E-mail: evagenas@academyofathens.gr} }\\
\newline
\\
{\normalsize \it $^1$ Department of Science, Payame Noor University, Bijar, Iran}
\\
{\small \it $^2$ Research Institute for Astronomy and Astrophysics of Maragha,}
\\
{\small \it P.O. Box 55134-441, Maragha, Iran}
\\
{\normalsize \it $^2$ Research Center for Astronomy \& Applied Mathematics,}
\\
{\normalsize \it Academy of Athens, Soranou Efessiou 4, GR-11527, Athens, Greece}}

\date{\small{}}

\maketitle

\begin{abstract}
Motivated by the recent interest in phantom fields as candidates for the dark energy component,
we investigate the consequences of the phantom field when is minimally coupled to gravity.
In particular, the necessary (but insufficient) conditions for the acceleration and superacceleration of the universe are obtained
when the non-minimal coupling term is taken into account. Furthermore, the necessary condition for the
cosmic acceleration is derived when the phantom field is non-minimally coupled to gravity and baryonic matter
is included.
\end{abstract}

\newpage

\section{Introduction}
Nowadays there is a general consensus that the universe is experiencing
an accelerated expansion. Recent observations from supernovae type Ia
\cite{1} associated with Large Scale Structure \cite{2}
and Cosmic Microwave Background anisotropies \cite{3} have
provided strong evidence for this cosmic acceleration.  To interpret
the cosmic acceleration, a so-called dark energy component has
been proposed. The nature as well as the microphysics of the dark energy
component are still ambiguous.
The simplest candidate for dark energy is a cosmological constant
with the equation of state parameter to be $\omega=-1$. Notwithstanding,
this scenario suffers from serious problems such as the huge fine tuning
and the coincidence problem \cite{4}. The field models that have
been widely discussed in the literature consider: (a) a canonical
scalar field named as quintessence \cite{quint}, (b) a phantom field, that is
a scalar field with a negative sign for the kinetic term
\cite{phant,phantBigRip}, or (c) the combination of quintessence and
phantom in a unified model named quintom \cite{quintom}.
Alternative approaches to explain the universe's late-time
acceleration include the  k-essence \cite{kessence}, the tachyon
\cite{tachyon}, the holographic dark energy
\cite{holoext} and many others.
Therefore, scalar fields play a crucial role in modern cosmology \cite{Elizalde:2008yf}.
\par\noindent
In the inflationary scenario scalar fields generate an exponential rate of
evolution of the universe as well as the density fluctuations due to
the vacuum energy. There are many theoretical evidences 
that suggest the incorporation of an explicit
non-minimal coupling (NMC) between a scalar field and gravity in the action
\cite{{far},{VF1}}. The nonzero non-minimal coupling arises from
the quantum corrections and it is also required by the renormalization
of the corresponding field theory. Amazingly, it has been proven
that the phantom divide line crossing of dark energy described by a
single minimally coupled scalar field with a general Lagrangian is
either dynamically unstable with respect to the cosmological perturbations,
or realized on the trajectories of zero measure \cite{vik}.\\
Furthermore, inflation is the theorized exponential expansion
of the universe at the end of the grand unification epoch, driven by
a negative-pressure vacuum energy density \cite{lin}. As a direct
consequence of this expansion, all of the observable universe
was originated in a small causally connected region. Inflation answers
the classic conundrum of the big bang cosmology: why does the
universe appear flat, homogeneous and isotropic in accordance with
the cosmological principle when one would expect, on the basis of
the physics of the big bang, a highly curved, heterogeneous
universe?
It is well known from the inflationary theory of the
primordial universe \cite{lid} that a scalar field $\phi$ slowly
rolling in a nearly flat section of its self-interaction potential
$V(\phi)$ can fuel a period of accelerated expansion of the
universe. The inflaton field $\phi$ satisfies the Klein-Gordon
equation in a curved spacetime and one needs to introduce, in general, a
non-minimal coupling term that accommodates the scalar field $\phi$ and the
Ricci curvature of spacetime $R$. The classical works on inflation
have neglected the coupling term (hereafter, this theory is called
ordinary inflation), as opposed to the generalized inflation which
has included the non-minimal coupling term.
\par
In the present paper we study the cosmic acceleration and thus the inflation when
a non-minimally coupled phantom scalar field is included. The remainder of the paper is as follows.
In Section 2, we present the field equations when the phantom field is present and
non-minimally coupled to the Ricci scalar. We have chosen a specific formulation
of the Einstein equation for which the corresponding energy-momentum tensor
is covariantly conserved. In Section 3, when the non-minimally coupling term is taken into consideration,
the necessary condition for the universe to accelerate, and thus inflate, is derived.
In Section 4, due to the presence of the non-minimal coupling term extra conditions compared to
the minimally coupled case, have to be satisfied in order the universe to superaccelerate.
In Section 5, apart from the non-minimally coupled phantom field, baryonic matter is included
and the necessary condition for the cosmic acceleration is obtained. Finally, in Section 6 a brief
description of the conclusions is given.
%
%
%
%
%
%
\section{Field equations for the phantom field}
\par\noindent
The action for a phantom field $\phi$ which is non-minimally coupled to gravity is given by
\be
S=\int d^{4}x \sqrt{-g}\left(\frac{R}{16\pi
G}+\frac{1}{2}\nabla^a \phi \nabla_a \phi-V(\phi)-\frac{\xi}{2}R
\phi^2 \right)
\label{1}
\ee
where $g$ is the determinant of the metric $g_{ab}$,
$G$ is the Newton's gravitational constant,
$R$ is the Ricci curvature of the spacetime, $V(\phi)$ is the
phantom field potential, and $-\frac{\xi}{2}R \phi^2$ depicts the non-minimal coupling of the
phantom field to the Ricci scalar.
\par\noindent
The dynamical equation for the phantom field is obtained from the variation of the
action (\ref{1}) with respect to the phantom field, i.e.
\be
\frac{\delta S}{\delta \phi}=0~,
\ee
and takes the form
\be
\Box\phi+\xi R
\phi+\frac{dV}{d\phi}=0~.
\label{2}
\ee
In the case that one wants to include matter fields $\psi_{m}$  in the system under consideration then
the action (\ref{1}) will be
\be
S=S_{EH}[g_{ab}]+S_{int}[g_{ab},\phi]+S_{\phi}[g_{ab},\phi]+S_{m}[g_{ab},\psi_{m}]
\label{3}
\ee
where
\bea
S_{EH}&=&\frac{1}{2\kappa}\int d^4x \sqrt{-g}R \nn\\
S_{int}&=&-\frac{\xi}{2}\int d^4x \sqrt{-g}R \phi^{2}\nn\\
S_{\phi}&=&-\int d^{4}x \sqrt{-g} \left(-\frac{1}{2}\nabla^a \phi \nabla_a \phi + V(\phi)\right)\nn
\label{4}
\eea
where $\kappa\equiv 8\pi G$ , $S_{EH}$ is the Einstein-Hilbert action
which is the purely gravitational part of action (\ref{1}),
$S_{int}$ is the part of the action that denotes the coupling between the gravitational and the phantom
field, $S_{\phi}$ is the action of phantom field, and $S_m$
describes the matter fields other than $\phi$. Thus, action (\ref{1}) can explicitly be written as
\be
S=\int d^4x \sqrt{-g}\left\{\left(\frac{1}{2\kappa}-\frac{\xi}{2}
\phi^{2}\right)R+\frac{1}{2}g^{ab}\nabla_a \phi \nabla_b
\phi-V(\phi)\right\}+S_{m}[g_{ab},\psi_{m}]~.
\label{5}
\ee
\par\noindent
The variation of Eq.(\ref{4}), or equivalently of Eq. (\ref{3}),
with respect to the metric field $g_{ab}$, yields the Einstein
equation slightly modified due to the presence of the coupling
\be
(1-\kappa\xi\phi^2)G_{ab}=\kappa\big(\tilde{T_{ab}}[\phi]+\tilde{T_{ab}}[\psi_m]\big)
\equiv \kappa\tilde{T_{ab}}^{(total)}
\label{6}
\ee
where
\be
\tilde{T_{ab}}[\phi]=
- \nabla_a \phi \nabla_b \phi
+ \frac{1}{2}g_{ab}\nabla^c \phi \nabla_c \phi
- g_{ab} V(\phi)
+ \xi\left(g_{ab}\Box \phi^{2}-\nabla_a \nabla_b \phi^2\right)
\label{7}
\ee
while the energy-momentum tensor of the matter fields is given by
\be
\tilde{T_{ab}}[\psi_{m}]=\frac{(-2)}{\sqrt{-g}}\frac{\delta
S_m[\psi_m, g_{cd}]}{\delta g^{ab}}~.
\label{8}
\ee
The modified Einstein equation (\ref{6}) can be rewrite as
\be
G_{ab}=\kappa\big(T_{ab}[\phi]+T_{ab}[\psi_{m},\phi]\big)
\equiv
\kappa T_{ab}^{(total)}
\label{9}
\ee
where
\be
T_{ab}[\phi]=\frac{1}{1-\kappa\xi\phi^2}\tilde{T_{ab}}[\phi]
\label{10}
\ee
\be
T_{ab}[\psi_{m},\phi]=\frac{1}{1-\kappa\xi\phi^2}\tilde{T_{ab}}[\psi_m]~.
\label{11}
\ee
At this point a couple of comments are in order.
First, the specific formulation described by Eq. (\ref{9})
of the modified Einstein equation (\ref{5}) adopted here is chosen due to the
fact that the corresponding energy-momentum tensor, i.e. $T_{ab}^{(total)}$, is covariantly conserved
as a result of the contracted Bianchi identities, i.e. $\nabla^{a}G_{ab}=0$, \cite{VF1}.
Second, one has to be very careful with the values of the coupling constant $\xi$.
In particular, when $\xi \leq 0$ the quantity
\be
C(\xi,\phi)=1-\kappa\xi\phi^{2}
\label{12}
\ee
is definitely positive and the possible formulations of the modified Einstein equation, mentioned in
the previous point, are all equivalent except from the covariant conservation of the
energy-momentum tensor that has to be investigated case by case. However, when $\xi > 0$
there is the possibility the phantom field to be equal to the roots of equation
$C(\xi,\phi)=0$, namely
\be
\phi_{\pm}=\pm \phi_{c}
\label{13}
\ee
where
\be
\phi_{c}=\frac{1}{\sqrt{\kappa\xi}}~.
\label{14}
\ee
In this case the corresponding energy-momentum tensor $T_{ab}^{(total)}$ diverges. Thus, the phantom field
has to avoid these values, i.e. $\phi\neq \phi_{\pm}$. Furthermore, when $|\phi|<\phi_{c}$,
or equivalently $\phi_{-}<\phi<\phi_{+}$, the quantity $C(\xi,\phi)$ is positive, i.e. $C(\xi,\phi)>0$,
while when $|\phi|>\phi_{c}$, or equivalently $\phi>\phi_{+}$ or $\phi<\phi_{-}$,
the quantity $C(\xi,\phi)$ is negative, i.e. $C(\xi,\phi)<0$. However, one has to take into account that
in one of the formulations of the modified Einstein equation $C(\xi,\phi)$ appears to be
inversely proportional to an effective Newton's gravitational constant. Therefore, $C(\xi,\phi)$ has
always to be positive and this is what we adopt here from now on. Finally, it is evident that
in the case of minimal coupling, i.e. $\xi=0$, $C(\xi,\phi)=1$.
%
\section{Phantom field and the acceleration of the universe}
\par\noindent
In this section, we are interested in the consequences of the presence
of the phantom field in the cosmic evolution of the universe and in particular during the
phase of cosmic acceleration, and hence of the inflation. For this purpose, we consider the
spatially flat Friedmann-Robertson-Walker (henceforth abbreviated to FRW) universe.
The line element of the FRW cosmological model  is given as
\be
ds^2=-dt^2+a^{2}(t)(dx^2+dy^2+dz^2)
\label{15}
\ee
where we have used the comoving coordinates $(t,x,y,z)$.
The Friedmann equations are given by
\be
H^{2}\equiv \left(\frac{\dot{a}}{a}\right)^{2}=\frac{\kappa}{3}\rho
\label{16}
\ee
\be
\frac{\ddot{a}}{a}=-\frac{\kappa}{6}(\rho+3p)
\label{17}
\ee
where $\rho$ and $p$ are the energy density and the pressure of the phantom field, respectively.
These two quantities are the diagonal components of the energy-momentum tensor $T_{ab}[\phi]$
and using (\ref{10}) one gets
\be
\rho=
\frac{1}{C(\xi,\phi)}
\Big[-\frac{1}{2}\dot{\phi}^{2}+V(\phi)+6\xi H\phi \dot{\phi}\Big]
\label{18}
\ee
\be
p=\frac{1}{C(\xi,\phi)}
\Big[\big(-\frac{1}{2} - 2\xi \big)\dot{\phi}^{2}-V(\phi)
- 2\xi \phi\ddot{\phi} - 4\xi H \phi\dot{\phi}\Big]~.
\label{19}
\ee
It is straightforward to rewrite the 1st and 2nd Friedmann equations, namely Eqs. (\ref{16}) and (\ref{17}) respectively,
in terms of the scale factor $a(t)$, the phantom field $\phi(t)$ and the
coupling constant $\xi$. Thus, substituting Eqs. (\ref{18}) and (\ref{19}) in Eqs. (\ref{16}) and (\ref{17}), the
Friedmann equations take the form
\be
H^{2}=\frac{\kappa}{3 C(\xi,\phi)}
\Big[-\frac{1}{2}\dot{\phi}^{2}+V(\phi)+6\xi H\phi \dot{\phi}\Big]
\label{20}
\ee
\be
\frac{\ddot{a}}{a}=\frac{\kappa}{3 C(\xi,\phi)}
\Big[ \left(1+3\xi\right)\dot{\phi}^{2}+V(\phi)+3\xi \phi \left(H\dot{\phi}+\ddot{\phi}\right)\Big]~.
\label{21}
\ee
Cosmic acceleration means $\ddot{a}(t)>0$ which in turn, using Eq. (\ref{17}), is translated to
the condition $\rho+3p< 0$. Employing Eq.(\ref{21}), this condition for acceleration is now written as
\be
\Big[ \left(1+3\xi\right)\dot{\phi}^{2}+V(\phi)+3\xi \phi \left(H\dot{\phi}+\ddot{\phi}\right)\Big] >0 ~.
\label{22}
\ee
At this point one needs to eliminate from the aforementioned condition the second time-derivative of
the phantom field. Therefore, one employs the Klein-Gordon equation (\ref{2}) which for the
spatially flat FRW model has the form
\be
\ddot{\phi}+3H \dot{\phi}-\xi R\phi-\frac{dV}{d\phi}=0~.
\label{23}
\ee
Using the above equation, namely Eq. (\ref{23}), as well as Eq. (\ref{18}), the condition for the cosmic acceleration,
i.e. Eq. (\ref{22}), becomes
\be
x \equiv C(\xi,\phi) \rho -
\Big[ \left(\frac{1}{2}+3\xi\right)\dot{\phi}^{2} + 2 V(\phi)
+ 3\xi^{2} R \phi^{2}  + 3\xi\phi\frac{dV}{d\phi}\Big] < 0~.
\label{24}
\ee
\par\noindent
Since the condition for the acceleration as given
by Eq. (\ref{24}) is quite cumbersome, one makes the first assumption
which is to adopt the weak energy condition, i.e. $\rho > 0$.
Therefore, the necessary (but not sufficient) condition for the cosmic acceleration
when the phantom field is non-minimally coupled becomes
\be
\Big[ \left(\frac{1}{2}+3\xi\right)\dot{\phi}^{2} + 2 V(\phi)
+ 3\xi^{2} R \phi^{2}  + 3\xi\phi\frac{dV}{d\phi}\Big] > 0~.
\label{25}
\ee
Now, we recall that  the Ricci scalar for the spatially flat FRW universe is of the form
\be
R=\frac{6\left(\dot{a}^{2}+a\ddot{a}\right)}{a^{2}}
\label{26}
\ee
which is a positive quantity if and only if the FRW universe is accelerating, i.e. $\ddot{a}>0$.
At this point, one has to make the second assumption and demands $\xi > 0$. Thus,
one obtains the necessary (but not sufficient) condition that
the phantom scalar potential has to satisfy for the cosmic acceleration, and thus for the inflation,
\be
V(\phi)+\frac{3}{2}\xi\phi \frac{dV}{d\phi} > 0 ~.
\label{27}
\ee
It is noteworthy that in the case of minimal coupling the necessary condition becomes
$V > 0$. It is obvious that in the case of the non-minimally coupled phantom field
it is ``easier" for the acceleration, and hence for the inflation, to be achieved.
However, it should be stressed that condition (\ref{27}) is not sufficient.
%
%
%
%
\section{Phantom field and the superacceleration}
\par\noindent
It is known that the spatially flat FRW universe superaccelerates, i.e. $\dot{H} > 0$,
when a phantom field is minimally coupled. In particular, when the phantom field
is minimally coupled the two Friedmann equations are written as
\be
H^{2}=\frac{\kappa}{3}
\Big[-\frac{1}{2}\dot{\phi}^{2}+V(\phi)\Big]
\label{28}
\ee
\be
\frac{\ddot{a}}{a}=\dot{H}+H^{2}=\frac{\kappa}{3}
\Big[ \dot{\phi}^{2}+V(\phi))\Big]~.
\label{29}
\ee
As it has already been mentioned, acceleration means $(\ddot{a}/a
)=\dot{H}+H^{2} > 0$. However, if Eqs. (\ref{28}) and (\ref{29})
are combined, one obtains
\be
\dot{H}=\frac{\kappa}{2}\dot{\phi}^{2}~,
\label{30}
\ee
thus $\dot{H} > 0$ and we say that the universe superaccelerates when $\dot{\phi}^{2} \neq 0$, or equivalently
that the phantom field is a form of superquintessence.
It is therefore interesting to see if the phantom field continues to be a form
of superquintessence when is non-minimally coupled.
\par\noindent
From Eq. (\ref{21}), one obtains
\be
\dot{H}+H^{2}=\frac{\kappa}{3 C(\xi,\phi)}
\Big[ \left(1+3\xi\right)\dot{\phi}^{2}+V(\phi)+3\xi \phi \left(H\dot{\phi}+\ddot{\phi}\right)\Big]
\label{31}
\ee
and if Eq. (\ref{20}) is employed, one obtains the equivalent of Eq. (\ref{30}) for the case
of non-minimal coupling, namely
\be
\dot{H}=\frac{\kappa}{2 C(\xi,\phi)}
\Big[ \left(1+2\xi\right)\dot{\phi}^{2} + 2\xi \phi \left(\ddot{\phi}-H\dot{\phi}\right)\Big]~.
\label{32}
\ee
It is evident that when the phantom field is non-minimally coupled the
spatially flat FRW universe does not superaccelerates if extra conditions are not satisfied.
%
%
\section{Phantom field,  matter and the acceleration}
\par\noindent
It is of interest to add baryonic (ordinary) matter which is pressureless, i.e. $p=0$,
with energy density $\rho_m\propto a^{-3}$, to the content of the universe.
The total energy density of the universe is now given as
\bea
\rho &=& \frac{1}{C(\xi,\phi)}
\Big[\rho_{m}-\frac{1}{2}\dot{\phi}^{2} + V(\phi)+6\xi H\phi\dot{\phi}\Big]\nn \\
& =& \frac{\rho_m}{C(\xi,\phi)}+\rho_{\phi}~.
\label{33}
\eea
Following the analysis of Section 3, the necessary and sufficient condition for the acceleration of the
universe (and thus for the inflation), i.e. $\rho+3p< 0$, takes the form
\be
\frac{\rho_m}{2}-(1+3\xi)\dot{\phi}^{2}-V(\phi)+6\xi H
\phi\dot{\phi} - 3\xi^{2}\phi^{2}R - 3\xi\phi\frac{dV}{d\phi}< 0 ~.
\label{34}
\ee
Implementing Eq. (\ref{33}), the necessary and sufficient condition takes the form
\be
y=-\frac{\rho_m}{2} + C(\xi,\phi)\rho -(\frac{1}{2}+3\xi)\dot{\phi}^{2}
- 2V - 3\xi^{2}R\phi^{2} - 3\xi\phi\frac{dV}{d\phi}<0
\label{35}
\ee
or, equivalently
\be
y=\frac{\rho_m}{2} + C(\xi,\phi)\rho_{\phi} -(\frac{1}{2}+3\xi)\dot{\phi}^{2}
- 2V - 3\xi^{2}R\phi^{2} - 3\xi\phi\frac{dV}{d\phi}<0 ~.
\label{36}
\ee
If one makes again the same assumptions, namely that the energy densities $\rho_{m}$ and
$\rho_{\phi}$ are positive and that the coupling constant takes only positive values, then
the necessary condition for the acceleration of the universe to be achieved is as before
\be
V(\phi)+\frac{3}{2}\xi\phi \frac{dV}{d\phi} > 0 ~.
\label{37}
\ee
%
%
%
\section{Conclusions}
\par\noindent
It is widely believed that our universe is in a phase of accelerating expansion.
This belief is mainly supported by observational data that has shown up in the last ten years.
One of the ways to interpret the acceleration is through the presence of a dark energy component
whose nature and microphysics are still lacking.
Scalar fields are a key player in this scenario. In particular, phantom scalar field
is one of the main candidates for dark energy.
\par\noindent
In this work we investigate the consequences of the presence of the phantom field
when it is non-minimally coupled to gravity.
The necessary condition for the acceleration, and so for the inflation, of the universe is derived.
It is easily seen that the acceleration is ``easier" achieved when the non-minimal
coupling term is present. However, this is not the case for the superacceleration.
It is known that when the phantom field is minimally coupled the universe superaccelarates.
Nevertheless, the necessary condition in the presence of the non-minimal coupling term is
quite complicated and extra conditions should be satisfied in order to get superaccelaration.
Furthermore, we consider the case where baryonic matter is present. The necessary condition
for the cosmic acceleration is the same with the one in the case where there is no matter at all.
Finally, it should be stressed that all aforesaid conditions are necessary but not sufficient
and thus one has to be very careful when handling them since one is not accredited to claim that
it is easier to get cosmic acceleration and thus inflation when the non-minimal
coupling of the phantom field is employed.
%
%
%
\section{Acknowledgments}
The work of M. R. Setare has been supported financially by
Research Institute for Astronomy and Astrophysics of Maragha, Iran.

\end{document}